\documentclass[superscriptaddress,letterpaper, amssymb,preprint, amsmath,prl 12pt]{revtex4-1}
 \usepackage[justification=raggedright,singlelinecheck=false,labelfont=bf,labelsep=period]{caption,subcaption}
\usepackage{braket}
\usepackage{siunitx}
\usepackage{comment}
\usepackage{nameref}
\usepackage[version=4]{mhchem}

\usepackage{natbib}
 
\usepackage{scalerel}
\usepackage{tikz}
\usetikzlibrary{svg.path}
\definecolor{orcidlogocol}{HTML}{A6CE39}
\tikzset{
  orcidlogo/.pic={
    \fill[orcidlogocol] svg{M256,128c0,70.7-57.3,128-128,128C57.3,256,0,198.7,0,128C0,57.3,57.3,0,128,0C198.7,0,256,57.3,256,128z};
    \fill[white] svg{M86.3,186.2H70.9V79.1h15.4v48.4V186.2z}
                 svg{M108.9,79.1h41.6c39.6,0,57,28.3,57,53.6c0,27.5-21.5,53.6-56.8,53.6h-41.8V79.1z M124.3,172.4h24.5c34.9,0,42.9-26.5,42.9-39.7c0-21.5-13.7-39.7-43.7-39.7h-23.7V172.4z}
                 svg{M88.7,56.8c0,5.5-4.5,10.1-10.1,10.1c-5.6,0-10.1-4.6-10.1-10.1c0-5.6,4.5-10.1,10.1-10.1C84.2,46.7,88.7,51.3,88.7,56.8z};}}
\newcommand\orcidicon[1]{\href{https://orcid.org/#1}{\mbox{\scalerel*{
\begin{tikzpicture}[yscale=-1,transform shape]
\pic{orcidlogo};
\end{tikzpicture}
}{|}}}}

\usepackage{etoolbox}

\makeatletter
\pretocmd\frontmatter@thefootnote{\color{blue}}{}{}

\usepackage{dcolumn} 
\usepackage{titlesec}
\titlespacing*{\section}
{0pt}{4ex}{1.3ex}
\titlespacing*{\subsection}
{0pt}{4ex}{0ex}

\titleformat{\section}
  {\bfseries\MakeUppercase}{\thesection\raggedright}{1em}{}
  \titleformat{\subsection}
 {\bfseries}{\thesubsection\raggedright}{1em}{}

\usepackage{xstring}
\usepackage{xspace}

\usepackage[utf8]{inputenc}
\usepackage[english]{babel}
\usepackage{natbib}
\usepackage{gensymb}
\usepackage{mathtools}

\usepackage{float}
\usepackage{graphicx}

\usepackage{url}
\usepackage[utf8]{inputenc}
\usepackage[T1]{fontenc}
\usepackage{textcomp}
\usepackage{gensymb}

\usepackage[colorlinks,citecolor=red,urlcolor=blue,bookmarks=false,hypertexnames=true]{hyperref}

\begin{document}

\title{Isotropic plasma-thermal atomic layer etching of aluminum nitride using SF\textsubscript{6} plasma and Al(CH\textsubscript{3})\textsubscript{3} }

\author{Haozhe Wang~\orcidicon{0000-0001-5123-1077} }

\affiliation{%
Division of Engineering and Applied Science, California Institute of Technology, Pasadena, California 91125, USA
}%
\author{Azmain Hossain}

\affiliation{%
Division of Engineering and Applied Science, California Institute of Technology, Pasadena, California 91125, USA
}%
\author{David Catherall}

\affiliation{%
Division of Engineering and Applied Science, California Institute of Technology, Pasadena, California 91125, USA
}%

\author{Austin J. Minnich~\orcidicon{0000-0002-9671-9540} }
 \email{aminnich@caltech.edu}
\affiliation{%
Division of Engineering and Applied Science, California Institute of Technology, Pasadena, California 91125, USA
}%      

\date{\today}
               
\begin{abstract}

We report the isotropic plasma atomic layer etching (ALE) of aluminum nitride using sequential exposures of SF\textsubscript{6} plasma and trimethylaluminum (Al(CH\textsubscript{3})\textsubscript{3}, TMA). ALE was observed at temperatures greater than 200 $^\circ$C, with a maximum etch rate of 1.9 \AA/cycle observed at 300 $^\circ$C as measured using ex-situ ellipsometry. After ALE, the etched surface was found to contain a lower concentration of oxygen compared to the original surface and exhibited a $\sim 35$\% decrease in surface roughness. These findings have relevance for applications of AlN in nonlinear photonics and wide bandgap semiconductor devices.

\end{abstract}
\maketitle

% \section*{Main text}
% high thermal conductivity \cite{jackson1997}, 

Aluminum nitride (AlN) is a III–V semiconductor of interest in  photonics and electronics owing to its simultaneous strong second-order ($\chi^{(2)}$) and third-order ($\chi^{(3)}$) optical nonlinearities , wide bandgap ($> 6$ eV), and high dielectric constant ($\sim 8.9$) \cite{xiong2012,lu2018, dong2019}. AlN has the lowest optical loss among III-V group materials over a range of wavelengths from the the ultraviolet to the mid-infrared \cite{lu2018,lin2014} and is under investigation for applications in ultraviolet light emitting diodes \cite{taniyasu2006, zhao2015} and optical quantum circuits \cite{jang2022,wan2020l}. High quality factors ($\gtrsim 10^6$) have been achieved in AlN microring resonators \cite{pernice2012high,sun2019} which are fundamental components of on-chip frequency combs and second-harmonic generation elements \cite{liu2021, pernice2012second, Hickstein2017}. AlN also finds potential applications as a low-leakage gate dielectric \cite{oikawa2015,zetterling1998,lee2004,adam2001} and has been employed in various thin film transistors \cite{zan2006,de2008,besleaga2016}.

For these applications, limitations on figures of merit have been attributed to surface imperfections and microfabrication processes. For instance, the presence of a native oxide leads to unstable etch rates with dry etching processes \cite{buttari2003}. The surface and sidewall roughness of etched nanostructures is on the order of $1 - 4$ nm \cite{chen2021, pernice2012high, xiong2012}, leading to light scattering. Poor-quality surface material with a refractive index fluctuations may also lead to light scattering even on nominally smooth surfaces. These non-idealities result in waveguide loss and limit the quality factor of optical microresonators, which in turn affects the performance of on-chip photonic devices \cite{krasnokutska2018}.

Atomic layer etching (ALE) is an emerging subtractive nanofabrication process with the potential to address these limitations \cite{kanarik2015, george2020, sang2020, knoops2019}. In isotropic ALE, a surface is etched using sequential reactions consisting of an initial surface modification followed by volatilization of the modified surface layer. Early development of ALE focused on directional etching using bombardment of a suitably prepared surface with low energy ions or neutral atoms \cite{yoder1988,sakaue1990,horiike1990}. Recently, isotropic ALE processes have been developed which enable isotropic etching with Angstrom-scale precision \cite{george2020}. Thermal and plasma isotropic ALE recipes are now available for various dielectrics and semiconductors including Al\textsubscript{2}O\textsubscript{3} \cite{lee2016,dumont2017al2O3,hennessy2017,zywotko2018,cano2019,chittock2020}, SiO\textsubscript{2} \cite{rahman2018,dumont2017siO2}, InGaAs \cite{lu2019}, and others \cite{george2020, fischer2021, fang2018}. Surface smoothing of etched surfaces using ALE has been reported for various materials including  Al\textsubscript{2}O\textsubscript{3} \cite{lee2015,zywotko2018}, amorphous carbon \cite{kanarik_2017}, and III-V semiconductors \cite{lu2019,ohba2017}. For AlN, isotropic ALE recipes have been reported using HF/Sn(acac)\textsubscript{2} and HF/BCl\textsubscript{3} \cite{johnson2016,cano2022}. However, identifying alternate reactants to HF vapor as well as realizing higher etch rates remain of interest for ALE of AlN.

% kanarik2018

Here, we report the atomic layer etching of AlN using  sequential exposures of SF\textsubscript{6} plasma and trimethylaluminum (Al(CH\textsubscript{3})\textsubscript{3}, TMA), achieving up to 1.9 \AA/cycle at 300 $^\circ$C. The necessity of both half-reactions for etching was established by verifying that no etching occurred with only  SF\textsubscript{6} plasma or TMA. The etched surface was characterized using atomic force microscopy and x-ray photoemission spectroscopy. The etched surface exhibited a decrease in roughness by $\sim 35$\% over a range of spatial frequencies after 50 cycles of ALE, and the ALE process was found to reduce the native oxide concentration at the surface. These improved surface characteristics highlight the potential of the process for applications in photonics and wide bandgap electronics.

\begin{figure}
 {\includegraphics[width = 450pt]{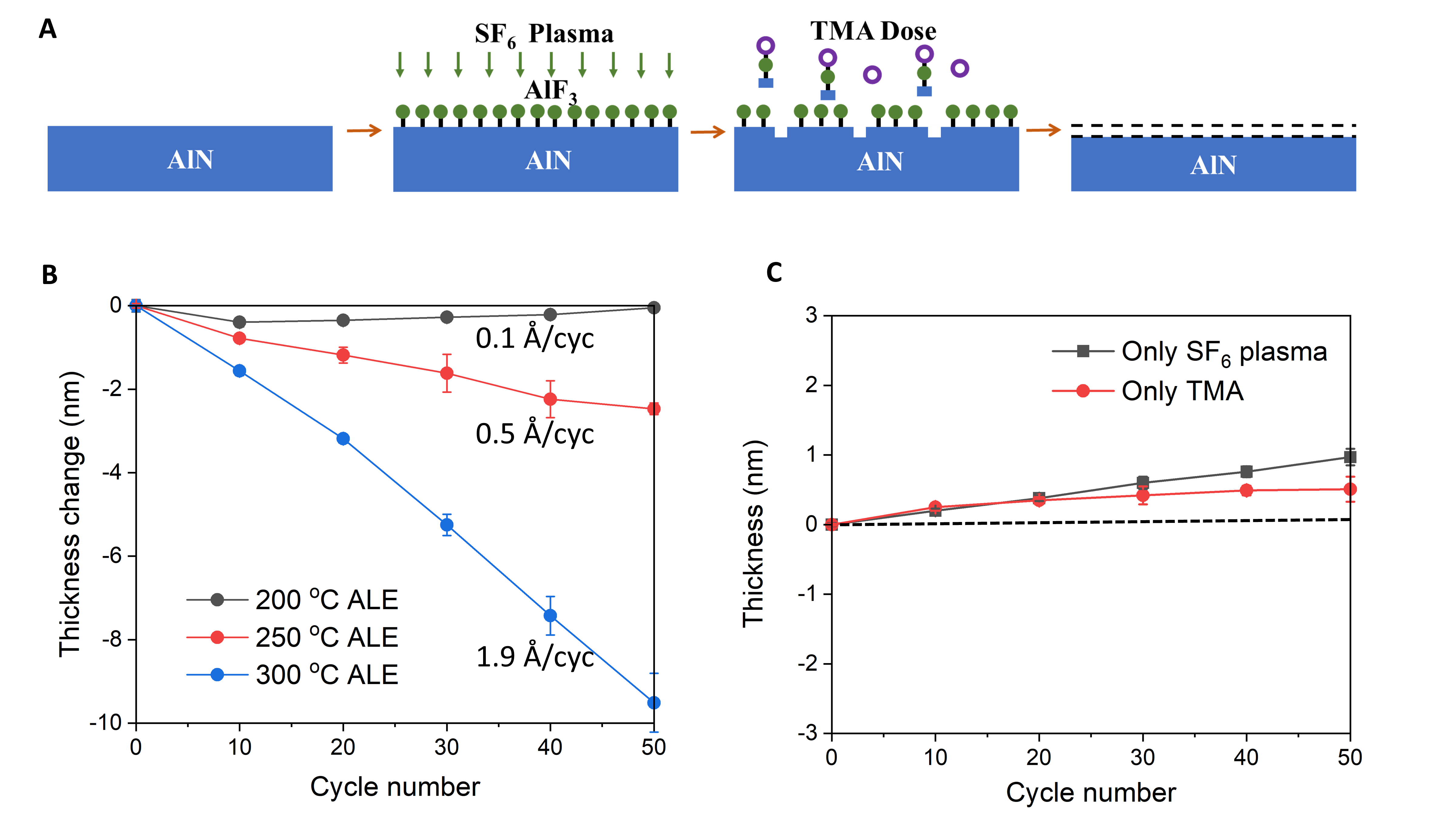}
    \phantomsubcaption\label{fig:demo}
    \phantomsubcaption\label{fig:etch rate}
    \phantomsubcaption\label{fig:self-limiting}}
\caption{
 (a) Isotropic plasma ALE of AlN. A low-power (100 W) SF\textsubscript{6} plasma containing F radicals (green dots) is first used to fluorinate the surface. TMA (purple circles) is introduced to perform ligand exchange with AlF\textsubscript{3}, yielding isotropic etching. (b) AlN thickness versus cycle number at different temperatures. The EPC is calculated based on the change in thickness after 50 cycles. (c) Thickness change versus cycle number for SF\textsubscript{6} plasma only and TMA-only recipes at  300$^\circ$C, confirming that etching requires both steps of the ALE process. Lines are guides to the eye. Error bars are as indicated or the size of the symbol.
}
\label{fig:ALE} 
\end{figure}

The sequence for the plasma-assisted thermal ALE process is illustrated in Fig.~\ref{fig:demo}. SF\textsubscript{6} plasma is first generated to fluorinate the surface using F radicals, producing AlF\textsubscript{3} on the AlN film surface. The excess gas phase reactants are then purged, and a TMA dose is introduced and held in the chamber. The TMA reacts with the AlF\textsubscript{3} in a ligand-exchange reaction, yielding volatile etching products \cite{clancey2019}. We hypothesize that the surface chemical reactions are similar to those reported for the isotropic plasma ALE of alumina using the same reactants \cite{chittock2020}; the specific reactions are a topic of future study.

This process was applied to AlN samples grown on Si (111) wafers by sputtering of an Al target gun with flow of 10 sccm nitrogen and 20 sccm Ar. The initial AlN films had a thickness of 280 \AA~as measured by spectroscopic ellipsometry (J. A. Woolam) at 65$^\circ$, 70$^\circ$ and 75$^\circ$ from 370 nm to 1000 nm. The samples were 12 × 12 mm\textsuperscript{2} chips. To determine the thickness change after a certain number of ALE cycles, 9 points were measured using ellipsometry on each AlN sample using a pre-programmed 10 × 10 mm\textsuperscript{2} square array with 5 mm spacing between points. Subsequently, the spectrum was fit by the Cauchy model to obtain the AlN thickness. The average thickness of a single sample was calculated using thicknesses measured and modeled by the 9 points. 

X-ray Photoemission Spectroscopy (XPS) analysis was performed using a Kratos Axis Ultra x-ray photoelectron spectrometer using a monochromatic Al K$\alpha$ source (Kratos Analytical). Depth profiling was performed using an Ar ion beam with a 60 s interval for each cycle. The total etch depth was measured by Dektak XT Stylus profilometer. The estimated etch depth for each cycle was then obtained by assuming the etching is uniform in all the cycles. The XPS data were analyzed in CASAXPS (Casa Software, Ltd.). As discussed in more detail below, the initial bulk composition of the films was found to be 55.9\% (Al) 41.3\% (N) 2.5\% (O), values which are consistent with other studies \cite{rosenberger2008,motamedi2014}.

Isotropic plasma ALE was performed using a FlexAL atomic layer deposition system (Oxford Instruments). The sample was placed on a 6-inch Si carrier wafer. The walls of the chamber were held at $\sim 150$ $^\circ$C to minimize reactant condensation. SF\textsubscript{6} plasma and trimethylaluminium (TMA) were used as the reactants. SF\textsubscript{6} plasma was struck with 30 sccm SF\textsubscript{6} and 150 sccm Ar mixing gas for 1 s and stabilized at 100 W power with 50 sccm SF\textsubscript{6} and 150 sccm Ar mixing gas for 2 s. The TMA was dosed with 100 sccm Ar carrier gas for 1 s. The precursor was then held in the chamber for 20 s without flowing gas or purging. The stage temperature was set at 200, 250, and 300 $^\circ$C.

Fig.~\ref{fig:etch rate} shows the AlN film thickness change versus cycle number for different process temperatures with other parameters fixed. At 200 $^\circ$C, we observe an initial etch in the first 10 cycles followed by a subsequent thickness increase, with an overall etch rate of 0.1 {\AA}/cycle over 50 cycles. At 250 $^\circ$C, a monotonic trend of thickness change is observed with an EPC of 0.5 {\AA}/cycle. The maximum EPC of 1.9 {\AA}/cycle was achieved at 300 $^\circ$C. The EPC increase with temperature is generally consistent with prior thermal ALE studies \cite{george2020} and those using the isotropic plasma ALE approach employed here \cite{chittock2020}.

To confirm that both half-reactions are required for etching, in Fig.~\ref{fig:self-limiting} we show the measured thickness change versus cycle number for SF\textsubscript{6}-plasma-only and TMA-only processes at 300$^\circ$C. No etching is observed with either recipe. These results support the proposed atomic layer etching mechanism requiring both half-reactions.

\begin{figure}
    \centering
    
{\includegraphics[width = 1\textwidth]{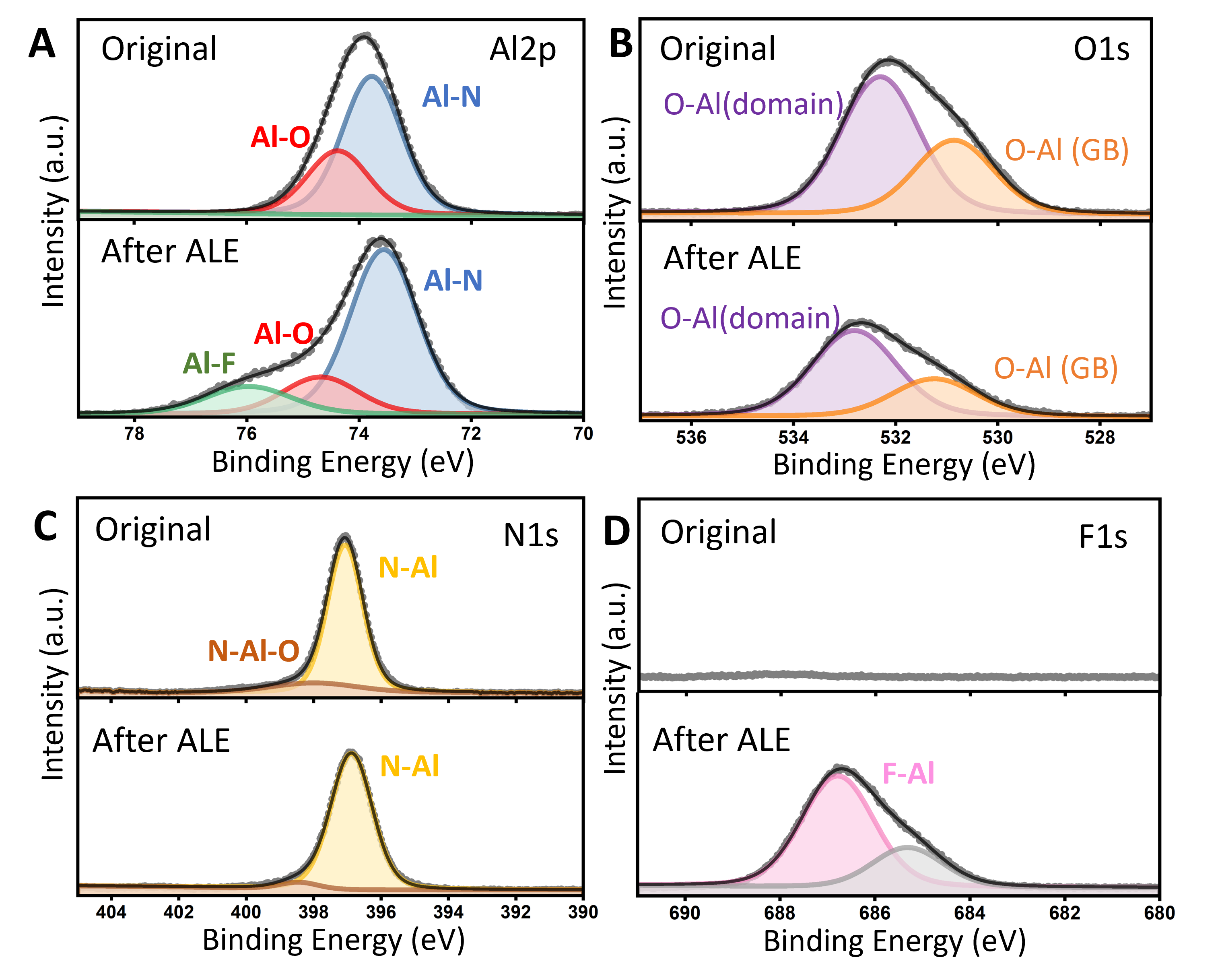}
    \phantomsubcaption\label{fig:xps_Al2p}
    \phantomsubcaption\label{fig:xps_O1s}
    \phantomsubcaption\label{fig:xps_N1s}
    \phantomsubcaption\label{fig:xps_F1s}
}
\caption{
AlN surface chemistry as characterized by XPS. XPS spectra of (a) Al2p, (b) N1s, (c) O1s and (d) F1s spectra before (up) and after (down) ALE. GB: grain boundary. The measured and fit spectra are shown as the gray dots and black line, respectively. The x-axis is the binding energy and y-axis is intensity in arbitrary units.
}

\label{fig:xps} 
\end{figure}

The chemical composition of original AlN thin films was characterized by XPS. In Fig.~\ref{fig:xps}, the XPS core levels for the Al2p, N1s, O1s, and F1s peaks are shown. We adopt the subpeak assignment from Ref. \cite{rosenberger2008} unless specified otherwise. For the Al2p spectrum (Fig.~\ref{fig:xps_Al2p}) of the original AlN thin film, we observe two subpeaks at 73.7 eV and 74.4 eV, assigned to Al–N  and Al–O bonds, respectively. The O1s energy profile for AlN before ALE is shown in Fig.~\ref{fig:xps_O1s}, and two subpeaks separated by 1.4 eV can be identified. We assign these two subpeaks to the O-Al bond in the bulk and grain boundaries, respectively. The N1s core level spectrum (Fig.~\ref{fig:xps_N1s}) has a primary N-Al subpeak at 396.8 eV and a secondary N-Al-O subpeak at 398.0 eV.

\begin{figure}
    \centering
    
{\includegraphics[width = 1\textwidth]{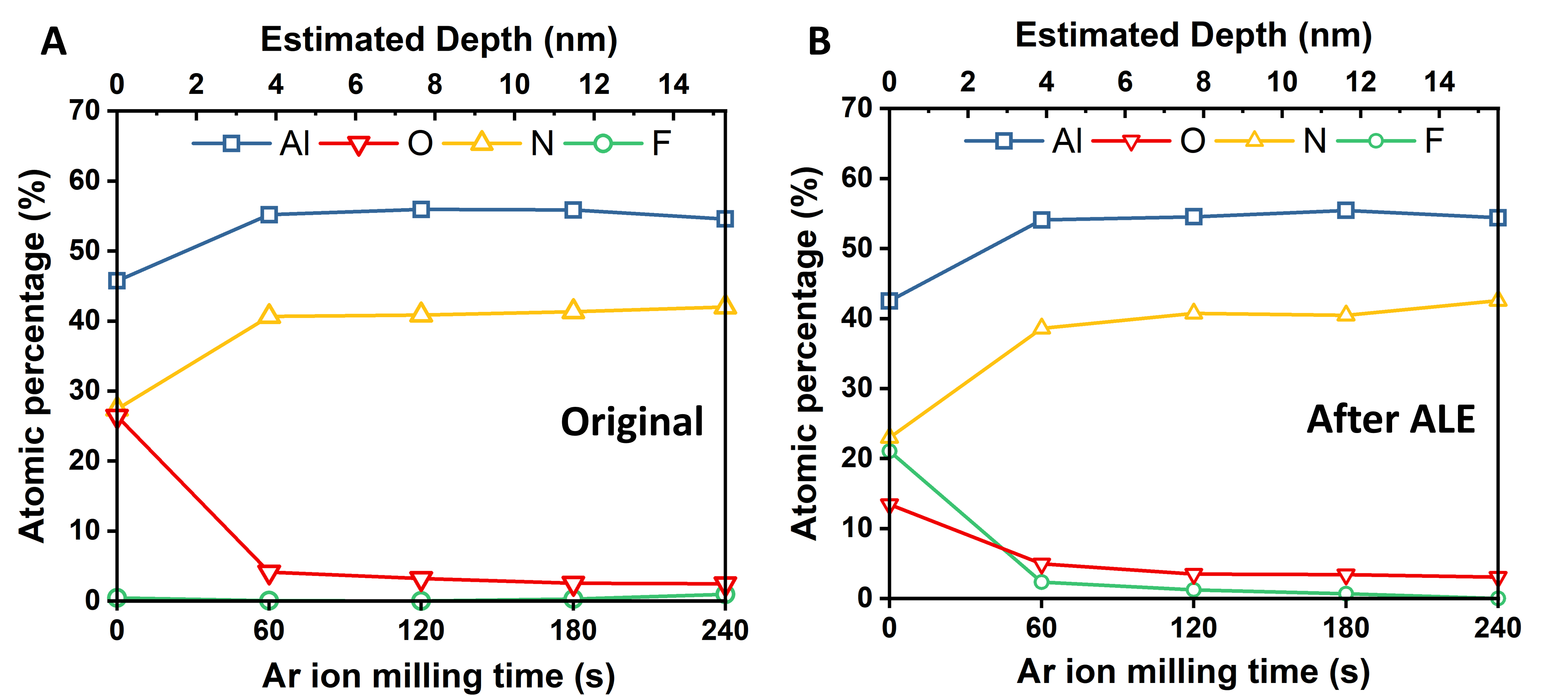}
    \phantomsubcaption\label{fig:depth_ori}
    \phantomsubcaption\label{fig:depth_ale}
}
\caption{Atomic concentration of Al, O, N, F versus XPS Ar ion milling time for (a) original and (b) ALE-treated AlN thin films. }
\label{fig:xps} 
\end{figure}

We performed XPS depth profiling to determine the atomic concentration in the surface and bulk. As shown in Fig.~\ref{fig:depth_ori}, the atomic concentrations at the surface of original AlN film are 45.7\% (Al), 27.4\% (N), and 26.4\% (O). Below $\sim 4$ nm (after 60 s Ar ion beam exposure), the atomic concentrations plateau to their bulk values of 55.2\% (Al) 40.6\% (N) 4.13\% (O). The higher oxygen concentration at the surface is consistent with the presence of a native oxide \cite{rosenberger2008,motamedi2014}.

We now examine the composition after ALE. In Fig.~\ref{fig:xps_F1s}, we observe the appearance of an F1s peak after ALE, indicating the presence of residual fluorine in the surface. The F1s peak has primary subpeaks at 686.8 eV, assigned to Al-F which is consistent with prior studies on fluorinated AlN \cite{watanabe2005}. Prior work also reported an additional subpeak with weaker intensity; \cite{hennessy2016,moreno2018,fischer2017plasma} this subpeak is also observed here at  685.2 eV. This subpeak lacks a known assignment. For the Al2p spectrum after ALE shown in Fig.~\ref{fig:xps_Al2p}, an additional subpeak at 76.0 eV is observed and assigned to the Al-F bond \cite{watanabe2005}. The O1s energy profile after ALE is shown in Fig.~\ref{fig:xps_O1s} with same O-Al subpeaks for bonds in the domain and at grain boundaries. For N1s, the position of secondary peak is shifted to 398.4 eV but the intensity is $<5$\% of the total N1s intensity. To the best of our knowledge, a bond assignment for this subpeak is not available.

The XPS depth profile of the ALE-treated AlN sample is shown in Fig.~\ref{fig:depth_ale}. It is notable that the surface oxygen concentration decreased to 13.5\%, indicating a lesser presence of native oxide compared to the original film. The fluorine concentration is found to be 21\% at the surface. After 60 s Ar ion milling, the fluorine concentration decreases to 2\% and the fractions of Al and N are within 95\% of those in the original film. This observation confirms that the alteration to the chemical composition of the film from the SF\textsubscript{6} plasma is confined to within a few nanometers of the surface, consistent with the findings of other works involving the interactions of fluorine plasmas with dielectric films \cite{fischer2017plasma}.

% , and the present ALE process does not alter the bulk film composition.

\begin{figure}
    \centering
{\includegraphics[width = 450pt]{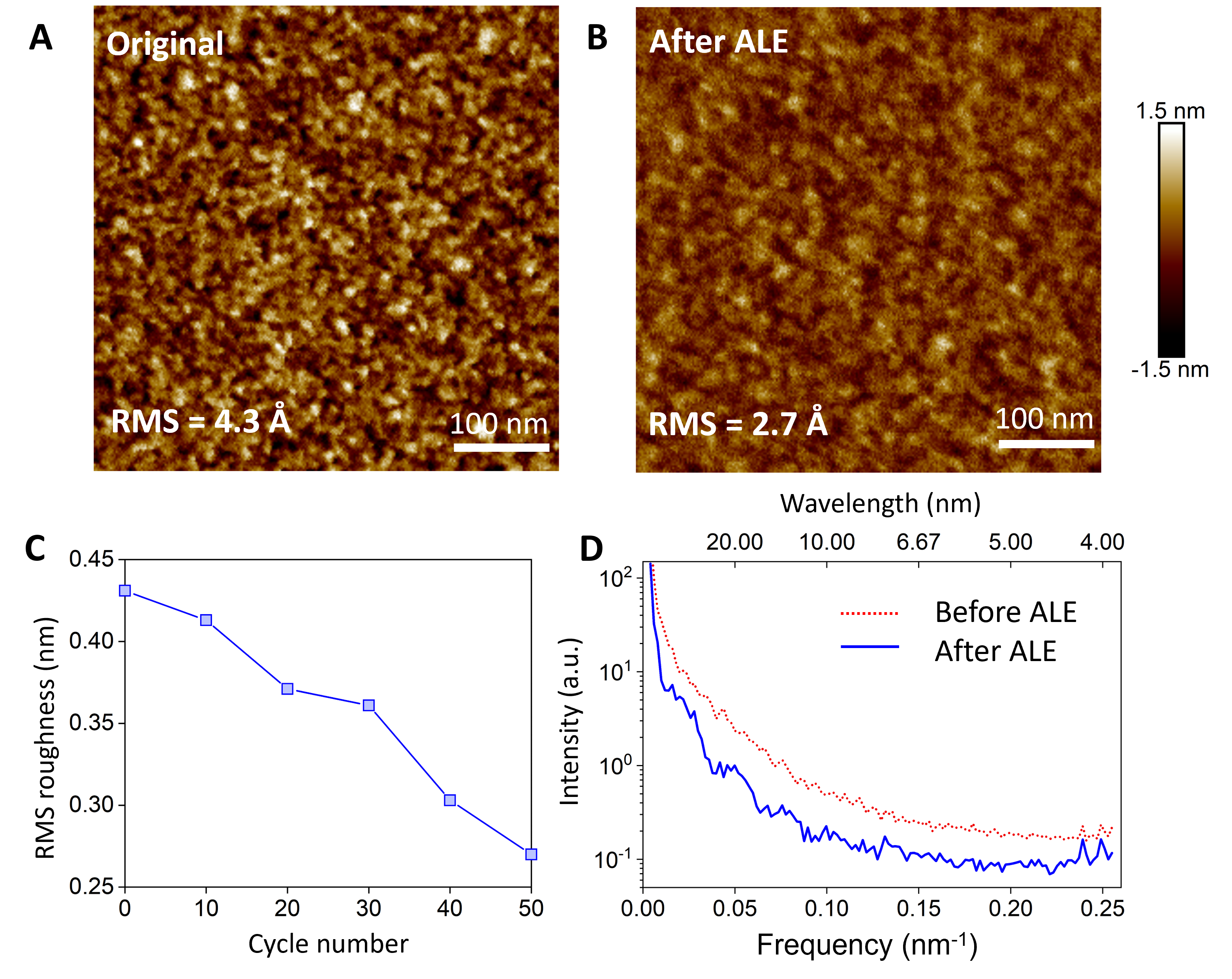}
    \phantomsubcaption\label{fig:AFM_ori}
    \phantomsubcaption\label{fig:AFM_ale}
    \phantomsubcaption\label{fig:roughness vs cycle}
    \phantomsubcaption\label{fig:gsd}}
\caption{
Smoothing effect of isotropic plasma ALE. AFM image of (a)  original surface and (b) after 50 cycles of ALE at 300 $^\circ$C. The root mean square (RMS) roughness is calculated and labeled in the images. (c) Surface roughness versus ALE cycle number, indicating a monotonic decrease in roughness with cycle number. Lines are guides to the eye. (d) Surface PSD (arbitrary units) versus spatial frequency (bottom axis) and wavelength (top axis) for the original sample (red dotted line) and the etched sample (blue solid line). ALE uniformly decreases surface roughness over a range of spatial frequencies.}
\label{fig:smoothing} 
\end{figure}

We next characterized the surface roughness of the film before and after ALE using Atomic Force Microscopy (AFM). Figs.~\ref{fig:AFM_ori} and \ref{fig:AFM_ale} show AFM images of the film before and after ALE at 300 $^\circ$C, respectively. Over an area of $0.5 \times 0.5$ $\mu$m\textsuperscript{2}, the RMS roughness decreased $\sim 35$\%, from 4.3 \AA~to 2.7 \AA, after 50 cycles of ALE. The RMS roughness versus cycle number is plotted in Fig.~\ref{fig:roughness vs cycle}. A monotonic decrease in surface roughness is observed with cycle number. This observation was reproduced on three separate regions on each sample.

The power spectral density (PSD) of the surface was computed using the measured AFM scans. The PSD provides a quantitative measure of the lateral distance over which the surface profile varies in terms of spatial frequencies \cite{elson1995,myers2021,dash2009,gong2016}. The PSD was calculated by removing tilt via linear plane-fit and subsequently performing a 1D-discrete Fourier transform over each row and column in the raw AFM data. The transformed data were then averaged along one dimension to produce single PSD curve. The PSD computed from Figs.~\ref{fig:AFM_ori} and ~\ref{fig:AFM_ale} are plotted in Fig.~\ref{fig:gsd}. A uniform decrease in roughness is observed over all spatial frequencies, indicating that both high and low spatial frequency components are smoothed by the ALE process.

We now discuss the characteristics of our isotropic plasma ALE process in context with thermal ALE processes for AlN and related materials. Thermal ALE of AlN has been reported previously using HF and Sn(acac)\textsubscript{2} \cite{johnson2016} and HF or XeF\textsubscript{2} and BCl\textsubscript{3} \cite{cano2022}. The maximum EPC reported for the latter process was 0.93 \AA~at a substrate temperature of 300 $^{\circ}$C \cite{cano2022}. Neglecting possible differences between stage and substrate temperatures, the present process achieves a nearly two-fold increase in EPC at a comparable temperature. The present process also enables similar EPCs as the thermal process at lower temperatures. Similar observations regarding higher etch rates at a given temperature using an isotropic plasma ALE process was reported in Ref.~\cite{chittock2020} for etching of alumina with the same reactants. These observations offer evidence for the benefits of isotropic plasma ALE processes compared with purely thermal methods. 

Our isotropic plasma ALE process may find potential applications in on-chip nonlinear and quantum photonics based on AlN, for which scattering by surface imperfections represents a primary limitation for various figures of merit.  Based on the measured PSD, our process decreases surface roughness of features with periods up to tens of nanometers. This smoothing capability may enable the reduction in sidewall roughness due to reactive ion etching as well as lithographic roughness transferred from the resist to the film, which would in turn reduce optical losses. 

In summary, we reported an isotropic plasma ALE process for AlN using sequential SF\textsubscript{6} plasma and Al(CH\textsubscript{3})\textsubscript{3} exposures. The etch rate reaches a maximum of 1.9 {\AA}/cycle at 300 $^{\circ}$C. We observe a smoothing effect from ALE, with a decrease in RMS roughness of $\sim 35$\% after 50 cycles. The surface oxygen content is significantly decreased after ALE, indicating that native oxides are largely removed by the process. We anticipate that the ability to engineer the surface of AlN films with enhanced precision using isotropic plasma ALE will facilitate applications of AlN in nonlinear photonics and electronics.

\section*{Acknowledgements}

This work was supported by Kavli Foundation and by the AFOSR under Grant Number FA9550-19-1-0321. The authors thank Nicholas Chittock, Guy DeRose, Harm Knoops, Kelly McKenzie, and Russ Renzas for useful discussions. We gratefully acknowledge the critical support and infrastructure provided for this work by The Kavli Nanoscience Institute and the Molecular Materials Research Center of the Beckman Institute at the California Institute of Technology.

% \subsection{Reference}

\bibliographystyle{is-unsrt}

\bibliography{refs}

\end{document}